\documentclass[10pt]{iopart}
\usepackage{graphicx}

\newcommand{\lcov }{\Delta_{\textrm{\tiny LB}}}

\newcommand{\Width }{\omega}
\newcommand{\anti }{\epsilon}
\newcommand{\erf }{\textrm{erf}}

\begin{document}
\jl{1}
\title[Signed zeros of Gaussian vector fields]{Signed zeros of Gaussian vector fields--density, correlation functions and curvature}
\author{Georg Foltin}
\address{Institut f\"ur Theoretische
Physik III, Heinrich--Heine--Universit\"at D\"usseldorf,
Universit\"atsstrasse
1, D--40225 D\"usseldorf, Germany}
\begin{abstract}
We calculate correlation functions of the (signed) density of zeros of Gaussian distributed vector fields. We are able to express correlation functions of arbitrary order through the curvature tensor of a certain abstract Riemann Cartan or Riemannian manifold. As an application, we discuss one and two-point functions. The zeros of a two-dimensional Gaussian vector field model the distribution of topological defects in the high temperature phase of two-dimensional systems with orientational degrees of freedom, such as superfluid films, thin superconductors and liquid crystals.
\end{abstract}

\section{Introduction}
\begin{figure}
\includegraphics[scale=0.75]{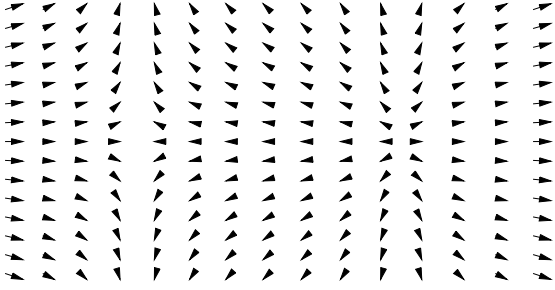}
\caption{\label{fig1}The planar vector field $(u_x,u_y)=(x^2-1,y)$
with a negative zero (left) and a positive zero (right).}
\end{figure}

Topological defects play an important role in two-dimensional physics, and are observed  in liquid crystal films \cite{Gen93}, two-dimensional crystals \cite{Hal78,Nel79}, superconducting films \cite{Bea79,Don79,Hal79}, films of superfluid helium \cite{Bis78}, arrays of Josephson contacts \cite{Abr82,Lob83}, nodal points of quantum and microwave billiards \cite{Seb99,Ber02,Bar02} and nodal points in optical waves \cite{Wei82,Fre94a}.
Common to these systems are orientational degrees of freedom with a global $O(2)$ symmetry.
The above systems are either in an orientationally ordered or disordered state.
Even in the ordered state, the orientational order is usually not perfect, but modified by the presence of massless spin waves. The corresponding correlation functions decay algebraically with distance (quasi long range order).
In the disordered state, orientational order is destroyed through  the spontaneous creation and subsequent unbinding of pairs of topological defects. The disordered phase is characterized by a finite density of these defects and exponentially decaying correlation functions.
A suitable order parameter for two-dimensional orientational order is a two component vector field $\bi{u}(\bi{r})=(u_1,u_2)(r_1,r_2)$.
Topological defects are points, where the amplitude of the director field $\bi{u}$ vanishes. Circling around a  defect, the phase of $\bi{u}$ adds up $\pm2\pi$ (or rarely multiples of $\pm2\pi$). We distinguish between positive and negative zeros (defects), where the sign of the
defects (sign of the phase jump) is equal to the sign of the Jacobian $\det(\partial_iu_j)$ right at the zero $\bi{u}=0$. Figure (\ref{fig1}) illustrates the two types of zeros.  
A crucial feature of the defects is the topological constraint on the number of positive zeros minus the number of negative zeros  on a closed surface  \cite{Bru82}
\begin{equation}
\label{topol}
\textrm{number of  positive zeros - number of  negative zeros}=2(1-\gamma)
\end{equation}
where $\gamma$ is the number of handles of the surface. A sphere has therefore at least two positive defects, whereas  a director field on a torus might be free of defects. Additional defects are created in pairs of opposite sign in order to obey the topological constraint (\ref{topol}).  
In this paper we consider the distribution of defects in the disordered (high temperature) phase, where the director field $\bi{u}$ has a Gaussian distribution. At first we
calculate general correlation functions of the (signed) defect density for arbitrary vector fields with a Gaussian distribution. As an application we study the mean 
defect density for certain distributions of $\bi{u}$ on curved manifolds.
Then we give a transparent derivation of the two-point correlation function for a particular class of vector fields. Finally we check the neutrality of the defect distribution.

Densities of zeros were discussed before in the context of nodal points of chaotic wavefunctions in microwave billiards \cite{Leb96,Sai01,Bar02},
random surfaces \cite{Lon57a, Lon57b,Lon60a,Lon60b,Lon60c,Car56,Adl81,Adl00,Tay01,Tay02}
and for the XY-model \cite{Hal81,Fol00}. Extensively studied are the optical dislocations (optical vortices) of ``chaotic'' optical wave fields \cite{Wei82,Fre94a, Ber78,Shv94a,Shv94b,Fre94b,Fre95a,Fre95b,Fre95c,Fre97b,Fre98,Kes98,Ber00,Ber01} which are well modelled by Gaussian distributed wave fields.
The distribution of zeros of Gaussian random polynomials and Gaussian random analytic functions was discussed in \cite{Bog92,Han96,Ble97,Ble00a,Ble00b,Ble00c}. Random polynomials and random analytic functions have only positive zeros\footnote{The real and imaginary part of an analytic function are harmonic functions, which have only saddle points as extremal points.}
and are therefore quite different from the systems considered here.

\section{Correlation functions}
We denote the positions of the zeros of a $d$-dimensional vector field $u_i$ in $d$-dimensional space by $\bi{r}_\alpha$ and the charge (type of the extremum) by $q_\alpha=\textrm{sign}\det\left(\partial_iu_j(\bi{r}_\alpha)\right)$. The charge density reads
$\rho(\bi{r})=\sum_\alpha q_\alpha\delta^d(\bi{r}-\bi{r}_\alpha)$
and can be written as
\begin{equation}
\label{chargedensity}
\rho(\bi{r})=\det\left(\partial_iu_j(\bi{r})\right)\,\delta^d\left(\bi{u}(\bi{r})\right).
\end{equation}
To prove this representation, consider a particular field $u_i$ with a zero at the origin and its expansion $u_i(\bi{r})= A_{ij}r_j+\ldots$ for small $\bi{r}$, where $A_{ij}=\partial_ju_i(\bi{r}=0)$ is the derivative of  $u_i$ at the origin and summation over double indices is implied. Then $\det(\partial_iu_j)=\det A+\ldots$ and $\delta^d(\bi{u})=\delta^d(A_{ij}r_j)=|\det A|^{-1}\delta^d(\bi{r})$. Combining both, we obtain
\begin{equation}
\det(\partial_iu_j)\delta^d(\bi{u})=\frac{\det A}{|\det A|}
\delta^d(\bi{r})=\pm\delta^d(\bi{r}).
\end{equation}
We consider now a Gaussian distributed vector field $u_i$ with correlations $\chi_{ij}(\bi{r},\bi{r}')=\left<u_i(\bi{r})u_j(\bi{r}')\right>$ and
provide a scheme to calculate arbitrary correlation functions $\left<\rho(\bi{r}_1)\ldots\rho(\bi{r}_f)\right>$
of the charge density $\rho$, where $\bi{r}_1,\ldots,\bi{r}_f$ are $f$ points and $\left<\ldots\right>$ denotes the average with respect to the Gaussian distribution of the field $u_i$.

We write a determinant as an integral over Grassmann variables $\theta_i, \eta_i$
\begin{equation}
\det A=\int(\rmd\eta\rmd\theta)^d\exp(\theta_i\eta_jA_{ij})
\end{equation}
(for an introduction to the calculus of Grassmann variables see \cite{Zin96}).
To regularize the theory it is helpful to replace the sharp delta function by a Gaussian with a small but finite
width $\Width$ and to use its Fourier representation
\begin{equation}\fl
\delta_\Width(\bi{x})=\frac{1}{(2\pi\Width)^{d/2}}\exp\left(-\frac{\bi{x}^2}{2\Width}\right)
=
(2\pi)^{-d}\int\rmd^dp\exp\left(-\frac{1}{2}\Width
\bi{p}^2-i\bi{p}\cdot\bi{x}\right).
\end{equation}
We obtain
\begin{equation}
\rho=\lim_{\Width\rightarrow 0}(2\pi)^{-d}\int\rmd^dp(\rmd\eta\rmd\theta)^d \exp\left(-\Width p^2/2+ \Psi(r,p,\theta,\eta)\right)
\end{equation}
where the field $\Psi$ depends on $2d$ bosonic coordinates $(r,p)$ and $2d$ fermionic variables $(\theta,\eta)$ and reads
\begin{equation}
\label{superfield}
\Psi(\bi{r},\bi{p},\bi{\theta},\bi{\eta})=ip_iu_i(\bi{r})+\theta_i\eta_j\partial_iu_j(\bi{r}).
\end{equation}
Since $\Psi$ is linear in $u_i$ and its derivatives, it has also a Gaussian distribution with correlations 
\begin{equation}\fl
\left<\Psi(x_A)\Psi(x_B)\right>=\left<\left(\theta_i^A\eta_j^A\partial_iu_j(\bi{r}_A)+ip_i^Au_i(\bi{r}_A)\right)
\left(\theta_k^B\eta_l^B\partial_ku_l(\bi{r}_B)+ip_k^Bu_k(\bi{r}_B)\right)\right>
\end{equation}
where  $x=(r,p,\theta,\eta)$.
The correlation function for finite width $\Width$ translates into
\begin{eqnarray}
\label{supercorr}
\left<\rho(\bi{r}_1)\ldots\rho(\bi{r}_f)\right>&=&
(2\pi)^{-fd}\int\rmd^{fd}p(\rmd\eta\rmd\theta)^{fd}\nonumber\\
&&\times\left<\exp\left(-(\Width/2)\sum_\alpha p^2_\alpha+\sum_\alpha
\Psi(x_\alpha)\right)\right>.
\end{eqnarray}
The Gaussian average can now be performed with the help of
\begin{equation}
\left<\exp\left(\sum_\alpha\Psi(x_\alpha)\right)\right>=\exp
\left(\frac{1}{2}\sum_{\alpha\beta} \left<\Psi(x_\alpha)\Psi(x_\beta)\right>\right)
\end{equation}
yielding (summation over double $i,j,k,l,m,n$ is implied, the partial derivatives act on the leftmost field only)
\begin{eqnarray}
\fl\left<\rho(\bi{r}_1)\ldots\rho(\bi{r}_f)\right>\nonumber\\
=\int (\rmd\eta\rmd\theta)^{fd}(2\pi)^{-fd}\int\rmd^{fd}p\exp\left(
\frac{1}{2}\sum_{\alpha\beta}\theta_i^\alpha\eta_j^\alpha\theta_k^\beta\eta_l^\beta
\left<\partial_iu_j(\bi{r}_\alpha)\partial_ku_l(\bi{r}_\beta)\right>\right.\nonumber\\
{}+i\sum_{\alpha\beta}\theta_i^\alpha\eta_j^\alpha p_k^\beta
\left<\partial_iu_j(\bi{r}_\alpha)u_k(\bi{r}_\beta)\right>
\nonumber\\
\left.-\frac{1}{2}\sum_{\alpha\beta}p_i^\alpha p_k^\beta\left(\Width\delta_{ik}\delta_{\alpha\beta}+\left<u_i(\bi{r}_\alpha)u_k(\bi{r}_\beta)\right>\right)\right).
\end{eqnarray}
After a straightforward (Gaussian) integration over the variables $p$ we find
\begin{eqnarray}
\lefteqn{\left<\rho(\bi{r}_1)\ldots\rho(\bi{r}_f)\right>=(2\pi)^{-fd/2}(\det g)^{-1/2}\int (\rmd\eta\rmd\theta)^{fd}}\nonumber\\
&&\times\exp\left(
\frac{1}{2}\sum_{\alpha\beta}\theta_i^\alpha\eta_j^\alpha\theta_k^\beta\eta_l^\beta
\Bigl(\left<\partial_iu_j(\bi{r}_\alpha)\partial_ku_l(\bi{r}_\beta)\right>\Bigr.\right.\nonumber\\
&&\left.\Bigl.
-\sum_{\mu\nu}\left<\partial_iu_j(\bi{r}_\alpha)u_m(\bi{r}_\mu)\right>g^{m\mu,n\nu}\left<u_n(\bi{r}_\nu)\partial_ku_l(\bi{r}_\beta)\right>\Bigr)\right)
\end{eqnarray}
where we have introduced the formal metric tensor
\begin{equation}
\label{metrictensor}
g_{i\alpha,j\beta}(\bi{r}_1,\ldots,\bi{r}_f)=\Width\delta_{ij}\delta_{\alpha\beta}+\left<u_i(\bi{r}_\alpha)u_j(\bi{r}_\beta)\right>
\end{equation}
and its inverse $g^{i\alpha,j\beta}=(g_{i\alpha,j\beta})^{-1}$, where a pair of a roman and a greek index $i\alpha$ is understood as a single composite index.
Now we interpret the tensor $g_{i\alpha,j\beta}$ as the metric tensor of a particular, $f\times d$-dimensional manifold. The manifold is parametrized by $f\times d$ internal coordinates $\sigma=(\bi{r}_1,\ldots,\bi{r}_f)$. It becomes singular in the limit $\Width\rightarrow 0$ whenever at least two points $\bi{r}_\alpha$ approach, since 
for two coinciding points two rows of the metric tensor become equal and 
$\det(g_{i\alpha,j\beta})=0$ therefore. These singular points show up as delta-like singularities in the correlation functions as shown in section \ref{sumrule}.
In addition to the metric tensor we introduce an affine connection
\begin{equation}
\Gamma^{k\gamma}_{j\beta,i\alpha}=\delta_{\alpha\beta}\left<\partial_iu_j(\bi{r}_\beta)u_m(\bi{r}_\mu)\right>g^{m\mu,k\gamma}.
\end{equation}
The affine connection  allows us to define a covariant derivative of a covariant vector field $V_{j\beta}$
\begin{equation}
D_{i\alpha}V_{j\beta}=\partial_{i\alpha}V_{j\beta}
-\Gamma^{k\gamma}_{j\beta,i\alpha}V_{k\gamma}
\end{equation}
where $\partial_{i\alpha}=\partial/\partial r^{i,\alpha}$.
The covariant derivative of a rank two tensor is\footnote{The covariant derivative of
higher-rank tensors is defined analogously, see \cite{Bru82}.}
\begin{equation}
D_{i\alpha}t_{j\beta,k\gamma}=\partial_{i\alpha} t_{j\beta,k\gamma}
-\Gamma^{m\mu}_{j\beta,i\alpha}t_{m\mu,k\gamma}
-\Gamma^{m\mu}_{k\gamma,i\alpha}t_{j\beta,m\mu}.
\end{equation}
A simple calculation shows the compatibility of the covariant derivative with the metric tensor $D_{i\alpha}g_{j\beta,k\gamma}=0$. The metric tensor together with the affine connection define a so-called Riemann-Cartan manifold, which is a generalization of a Riemannian manifold \cite{Bru82}. Unlike the latter, it has a non-zero torsion tensor $T^{k\gamma}_{i\alpha,j\beta}=\Gamma^{k\gamma}_{i\alpha,j\beta}
-\Gamma^{k\gamma}_{j\beta,i\alpha}$.
The commutator of two covariant derivatives defines the curvature tensor
\begin{equation}
\left(D_{i\alpha}D_{j\beta}
-D_{j\beta}D_{i\alpha}\right)V_{k\gamma}
=R_{k\gamma}{}^{l\lambda}{}_{i\alpha,j\beta}V_{l\lambda}+T^{m\mu}_{i\alpha,j\beta}
D_{m\mu}V_{k\gamma}.
\end{equation}
A straightforward calculation yields
\begin{eqnarray}
\label{riemann}
R_{k\gamma,l\lambda,i\alpha,j\beta}&=&
\delta_{\alpha\gamma}\delta_{\beta\lambda}
\left<\partial_iu_k(\bi{r}_\gamma)\partial_ju_l(\bi{r}_\lambda)
\right>\nonumber\\
&&{}-\delta_{\alpha\gamma}\delta_{\beta\lambda}
\left<\partial_iu_k(\bi{r}_\gamma)u_m(\bi{r}_\mu)\right>
g^{m\mu,n\nu}
\left<u_n(\bi{r}_\nu)\partial_ju_l(\bi{r}_\lambda)\right>
\nonumber\\
&&{}-(i\alpha)\leftrightarrow(j\beta)
\end{eqnarray}
where have used the metric tensor to transform a contravariant (upper) index into a covariant (lower) index $g_{i\alpha,j\beta} V^{j\beta}=V_{i\alpha}$.
The curvature tensor allows us to simplify the expression for the correlation function
\begin{eqnarray}
\label{mainresult}
\lefteqn{\left<\rho(\bi{r}_1)\ldots\rho(\bi{r}_f)\right>}\nonumber\\
&=&(2\pi)^{-fd/2}(\det g)^{-1/2}\int (\rmd\eta\rmd\theta)^{fd}\exp\left(
-\frac{1}{4}\theta_k^\gamma\theta_l^\lambda\eta_i^\alpha\eta_j^\beta
R_{k\gamma,l\lambda,i\alpha,j\beta}\right)\nonumber\\
&=&(2\pi)^{-fd/2}(\det g)^{1/2}\int (\rmd\eta\rmd\theta)^{fd}\exp\left(
-\frac{1}{4}\theta_k^\gamma\theta_l^\lambda\eta_i^\alpha\eta_j^\beta
R^{k\gamma}{}_{l\lambda}{}^{i\alpha}{}_{j\beta}\right).
\end{eqnarray}
It is quite remarkable that the correlation function is a purely geometrical object from the point of view of differential geometry---it depends on the metric tensor and the curvature tensor of a Riemann-Cartan manifold  with metric tensor
$g_{i\alpha,j\beta}=\Width\delta_{ij}\delta_{\alpha\beta}+\left<u_i(\bi{r}_\alpha)u_j(\bi{r}_\beta)\right>$ and affine connection $\Gamma^{k\gamma}_{j\beta,i\alpha}=\delta_{\alpha\beta}\left<\partial_iu_j(\bi{r}_\beta)u_m(\bi{r}_\mu)\right>g^{m\mu,k\gamma}$.
It should be noted that correlations of the number density \cite{Sai01,Ber00} (absolute value of the density) $|\rho|$  cannot be obtained by the above formalism.

In case of a vector field $u_i$ placed on a curved Riemannian  manifold with metric $G_{ij}(\sigma)$,
one has to refer to a covariant version of the density of extrema ($\sigma=(\sigma^1, \ldots,\sigma^d)$ are the internal coordinates to parametrize the manifold, $\nabla_i$ is the covariant derivative related to the metric $G_{ij}$)
\begin{eqnarray}
\rho(\sigma)&=&\det(G^{ij}\nabla_ju_k)\frac{1}{2\pi\Width}\exp\left(-\frac{1}{2\Width}G^{ij}u_iu_j\right)\nonumber\\
&\rightarrow&(\det G)^{-1/2}\det(\partial_iu_j)\delta^2(u_i)\textrm{ for }\Width\rightarrow 0
\end{eqnarray}
where $\nabla_iu_j=\partial_iu_j$ at the zeros of $u_i$.
The density  differs from the Euclidean version only by the factor $(\det G)^{-1/2}$.

\subsection{Gradient fields}
\label{sec-gradient}
A particular important class of vector field are the gradient fields $u_i=\partial_i\phi$
with a Gaussian distributed scalar field $\phi$. One might think of $\phi$ as a Gaussian random surface \cite{Lon57a,Lon57b} 
with local height $z=\phi(x,y)$.
The zeros of $\partial_i\phi$ are the extremal points of $\phi$. Extremal points with a positive signature are the maxima and minima of the function, whereas the saddle points of $\phi$ have a negative sign.
The scalar field $\phi$ should not be confused with the (polar) angle $\theta$ of a vector field $\bi{u}=u\,(\cos\theta,\sin\theta)$. The angular field $\theta$ is defined modulo $2\pi$ and has singularities (at the zeros of $\bi{u}$), whereas the field $\phi$ is nonsingular and single valued.  
The metric tensor (\ref{metrictensor}) for $\partial_i\phi$ reads
\begin{equation}
g_{i\alpha,j\beta}=\Width\delta_{ij}\delta_{\alpha\beta}+\left<\partial_i\phi(\bi{r}_\alpha)\partial_j
\phi(\bi{r}_\beta)\right>.
\end{equation}
(The latter metric tensor is also introduced in \cite{Adl00,Tay01,Tay02} in the context of excursion sets). The corresponding affine connection
\begin{equation}
\Gamma^{k\gamma}_{j\beta,i\alpha}=\delta_{\alpha\beta}\left<\partial_i\partial_j\phi(\bi{r}_\beta)\partial_m\phi(\bi{r}_\mu)\right>g^{m\mu,k\gamma}
\end{equation}
is symmetric in the lower indices, i.e. the torsion tensor $T^{k\gamma}_{i\alpha,j\beta}$ is zero. The vanishing torsion characterizes a Riemannian manifold,
where the affine connection is determined solely by the metric tensor $g$
\begin{equation}
\Gamma^{k\gamma}_{j\beta,i\alpha}=\frac{1}{2}\left(\partial_{j\beta}g_{m\mu,i\alpha}
+\partial_{i\alpha}g_{m\mu,j\beta}-\partial_{m\mu}g_{i\alpha,j\beta}\right).
\end{equation}
The Riemannian curvature tensor reads in this case
\begin{eqnarray}
\label{ex_riemann}
R_{k\gamma,l\lambda,i\alpha,j\beta}&=&
\delta_{\beta\lambda}\delta_{\alpha\gamma}
\left(\left<\partial_j\partial_l\phi(\bi{r}_\beta)\partial_i\partial_k\phi(\bi{r}_\alpha)\right>
\right.\nonumber\\
&&\left.{}
-\left<\partial_j\partial_l\phi(\bi{r}_\beta)\partial_m\phi(\bi{r}_\mu)\right>
g^{m\mu,n\nu}
\left<\partial_n\phi(\bi{r}_\nu)\partial_i\partial_k\phi(\bi{r}_\alpha)\right>
\right)\nonumber\\
&&{}-(i\alpha)\leftrightarrow(j\beta).
\end{eqnarray}
The defect density correlation function reads $\left<\rho(\bi{r}_1)\ldots\rho(\bi{r}_f)\right>=(2\pi)^{-fd/2}(\det g)^{1/2}\Gamma$, where the Grassmann integral
\begin{equation}
\Gamma=\int (\rmd\eta\rmd\theta)^{fd}\exp\left(
-\frac{1}{4}\theta_k^\gamma\theta_l^\lambda\eta_i^\alpha\eta_j^\beta
R^{k\gamma}{}_{l\lambda}{}^{i\alpha}{}_{j\beta}\right)
\end{equation}
is the so-called total curvature. It is zero for odd dimensions (here: dimension $=f\times d$). Its (covariant) integral over a (closed) surface is a topological invariant according to the theorem of Chern (1944) \cite{Che44,Bru82}. Indeed, the $f$-fold spatial integral over the manifold yields 
\begin{equation}
\prod_{\alpha=1}^{f}\int\rmd^dr_\alpha(\det G(\bi{r}_\alpha))^{1/2} \left<\rho(\bi{r}_1)\ldots\rho(\bi{r}_f)\right>=\left(2(1-\gamma)\right)^f.
\end{equation}
We will now specialize our main result (\ref{mainresult}).
At first we calculate one-point functions and two-point functions.

\section{Density of defects--the one point function}

\subsection{The high temperature phase of the XY-model}
We calculate the defect density of a XY-model, placed on an arbitrary curved surface, in the disordered (high-temperature) region \cite{Fol00}.
The Gaussian weight of a configuration $u_i$ in the high-temperature region is
\begin{equation}\fl
\label{htw}
P[u]\propto\exp\left(
-\frac{1}{2}\int\rmd A \left(\nabla_iu_j\nabla^iu^j+\left(\tau+(\eta+1) R/2\right)u_iu^i\right)\right).
\end{equation}
where $\rmd A=\sqrt{\det G_{ij}}\rmd^2\sigma$ is the invariant area element, $G_{ij}$ is the metric and  $\nabla_i$ is the covariant derivative of the surface \cite{Bru82} (a short introduction to the differential geometry of surfaces can also be found in \cite{Dav89}).
The mass $\tau$ determines the correlation length $\xi=\tau^{-1/2}$ of the vector field and, therefore, the distance from the critical point. In addition, we have introduced
a curvature dependent mass term as discussed in \cite{Fol00}.
The weight (\ref{htw}) is invariant under global $O(2)-$transformations
$u_i\rightarrow \cos(\alpha)u_i+\sin(\alpha)\anti_i{}^ju_j$, where $\anti_{ij}$ is the covariant antisymmetric unit tensor with respect to the surface.
The metric tensor reads $g_{ij}=\left<u_iu_j\right>=G_{ij}\left<u_mu^m\right>/2$, since $G_{ij}$ is the only symmetric rank two tensor invariant under $O(2)$.
The affine connection is
\begin{equation}
\Gamma_{ji}^k=\left<\partial_iu_ju_m\right>g^{mk}=\left<\nabla_iu_ju_m\right>
g^{mk}+\gamma_{ji}^k
\end{equation}
where $\gamma_{ji}^k$ is the affine connection of the underlying surface.
Due to the $O(2)$ invariance we can write \cite{Fol00}
\begin{equation}
\left<\nabla_i u_j u_m \right>= \frac{1}{2} G_{jm} \left<\nabla_i u_n u^n \right>
+ \frac{1}{2} \anti_{jm}
\anti^{np} \left<\nabla_i u_n u_p \right>
\end{equation}
and
\begin{equation}
\Gamma_{ji}^k=\gamma_{ji}^k+\anti_j{}^k\Omega_i+\delta_j^k\partial_i\log\left(\left<u^2\right>\right)/2
\end{equation}
where $\Omega_i=\anti^{np}\left<\nabla_iu_nu_p\right>/\left<u^2\right>$.
The gradient term does not contribute to the curvature tensor since it can be eliminated by a gauge transformation.
The curvature tensor reads ($R^s$ is the Riemann curvature tensor of the surface, $R$ is the scalar curvature of the surface)
\begin{equation}
R_k{}^l{}_{ij}=R^s{}_k{}^l{}_{ij}-(\nabla_i\Omega_j-\nabla_j\Omega_i)\anti_k{}^l
\end{equation}
and
\begin{equation}
R_{ijkl}=\frac{\left<u^2\right>}{2}\left(R/2-\anti^{mn}\nabla_m\Omega_n\right)\anti_{kl}\anti_{ij}.
\end{equation}
The Grassmann integral (\ref{twod}) is now trivial, and we obtain finally
\begin{equation}
\label{curvedenso2}
2\pi\rho=\left(R/2-\anti^{ij}\nabla_i\Omega_j\right).
\end{equation}
The topological constraint (\ref{topol}) for the total charge is easily verified since
\begin{equation}
2\pi\int\rmd A\rho=\int\rmd A(R/2-\anti^{ij}\nabla_i\Omega_j)=\int\rmd A\,R/2=4\pi(1-\gamma)
\end{equation}
due to the theorem of Stokes \cite{Dav89}.
This expression for the density of defects was already obtained by a direct calculation and further analysed in \cite{Fol00}.
It was shown that the signed density of zeros of the Gaussian XY-model with distribution (\ref{htw}) is equal (up to terms of order $\tau^{-2}$) to the charge density of the two-dimensional Coulomb gas in the Debye-H\"uckel approximation. This is not surprising, since both  approaches can be used to describe the high temperature region of the two-dimensional Coulomb gas \cite{Hal81}.  In the next section we show, that the extremal points of a Gaussian field $\phi$ behave in a similar manner.

\subsection{Incompressible and irrotational vector fields}
We calculate now the defect density $\left<\rho\right>$ for vector fields which are subject to an additional constraint, namely, the divergence or curl of the field must vanish.
A field with vanishing divergence (curl) can be expressed via a `potential' $\phi$ through $u_i=\anti_i{}^j\partial_j\phi$ or $u_i=\partial_i\phi$ respectively.
We plug this representation of $u_i$ into the weight (\ref{htw}) and obtain the statistical weight for the scalar field $\phi$
\begin{equation}
P[\phi]\propto\exp\left(
-\frac{1}{2}\int\rmd A((\lcov\phi)^2+(\tau+\eta R/2)\partial_i\phi\partial^i\phi)\right)
\end{equation}
where $\lcov=G^{ij}\nabla_i\partial_j$ is the covariant Laplace-Beltrami operator of the surface. 
The defects are the points where the vector field $u_i$ is zero, i.e. where the
potential $\phi$ has an extremal point. The charge of the defect is the sign of the derivate of $u_i$:
\begin{equation}
\textrm{sign} \det(\partial_iu_j)=\textrm{sign} \det(\partial_i\partial_j\phi)
\end{equation}
for both cases. The mean defect density is according to
theorem (\ref{mainresult})
\begin{eqnarray}
\label{twod}
\rho&=&(2\pi)^{-1}(\det G)^{-1/2}(\det g)^{-1/2}\int\rmd\eta_1\rmd\theta_1\rmd\eta_2\rmd\theta_2\nonumber\\
&&\times\exp\left(
\frac{1}{2}\theta_l\eta_j\theta_k\eta_i\left(\left<\partial_i\partial_k\phi\partial_j\partial_l\phi\right>-\left<\partial_i\partial_k\phi\partial_m\phi\right>g^{mn}\left<\partial_n\phi
\partial_j\partial_l\phi\right>\right)\right)\nonumber\\
&=&\frac{(\det g)^{1/2}}{2\pi(\det G)^{1/2}}\int\rmd\eta_1\rmd\theta_1\rmd\eta_2\rmd\theta_2
\exp\left(-\frac{1}{4}\theta_k\theta_l\eta_i\eta_j\,R^k{}_l{}^i{}_j\right)
\end{eqnarray}
where $g_{ij}=\left<\partial_i\phi\partial_j\phi\right>$. 
The Riemannian curvature tensor in two-dimensions has only one independent component--the scalar curvature $\tilde{R}=R^{kl}{}_{kl}$
\begin{eqnarray}
R_{klij}&=&\left<\partial_i\partial_k\phi\partial_j\partial_l\phi\right>-\left<\partial_i\partial_k\phi\partial_m\phi\right>g^{mn}\left<\partial_n\phi
\partial_j\partial_l\phi\right>-(i\leftrightarrow j)\nonumber\\
&=&\anti_{kl}\anti_{ij}\tilde{R}/2
\end{eqnarray}
where $\anti_{ij}$ is the covariant, antisymmetric unit tensor with respect to the metric $g_{ij}$: $\anti_{12}=\sqrt{\det g}, \anti_{21}=-\sqrt{\det g}$.
The Grassmann integral (\ref{twod}) is trivial, and we obtain finally
\begin{equation}
\label{curvedensity}
2\pi\rho\sqrt{\det G}=(\tilde{R}/2)\sqrt{\det g}.
\end{equation} 
The density of defects is according to equation (\ref{curvedensity}) the scalar curvature of an abstract surface (Riemannian manifold) with
metric $g_{ij}=\left<\partial_i\phi\partial_j\phi\right>$. Therefore, with the help of the Gauss-Bonnet theorem and of equation (\ref{curvedensity})
\begin{eqnarray}
\fl2\pi(\textnormal{number of positive defects}-\textnormal{number of negative defects})
&=&2\pi\int\rmd^2\sigma\sqrt{\det g}\rho\nonumber\\
\fl=\int\rmd^2\sigma\sqrt{\det g}\,\tilde{R}/2&=&4\pi(1-\gamma)
\end{eqnarray}
where $\gamma$ is the genus (number of handles) of both the abstract surface $g_{ij}$\footnote{
The number of handles is zero for a sphere and  one for a toroidal surface.} and the underlying surface $G_{ij}$, since the abstract surface is closely related to the underlying surface and especially inherits its topology.
The calculation of the curvature can be done in a large $\tau$-expansion (high-temperature expansion) for the special case $\eta=0$.
From general considerations, the first two terms of the $1/\tau$-expansion of
$g_{ij}$ read
\begin{equation}
\label{expansion}
g_{ij}=\left<\partial_i\phi\partial_j\phi\right>=AG_{ij}+B \frac{R}{\tau}G_{ij}+\mathcal{O}(\tau^{-2})
\end{equation}
with constants $A, B$ to be determined.  We multiply  equation (\ref{expansion})
with $G^{ij}$ and obtain ($R$ is the scalar curvature of the surface; the Laplace-Beltrami operator $\lcov$ refers to the surface as well) 
\begin{eqnarray}
G^{ij}g_{ij}&=&2A+2BR/\tau=\left<\partial^i\phi\partial_i\phi\right>=
\frac{1}{2}\lcov\left<\phi^2\right>-\left<\phi\lcov\phi\right>\nonumber\\
&=&\frac{1}{2\tau}\lcov\left(\left<\phi(\tau-\lcov)\phi\right>+\left<\phi\lcov\phi\right>\right)-\left<\phi\lcov\phi\right>\nonumber\\
&=&\frac{1}{4\pi}\log\left(
\frac{1}{a^2\tau}\right)-\frac{R}{12\pi\tau}+\mathcal{O}(\tau^{-2})
\end{eqnarray}
where $a$ is a short-distance cutoff. In \cite{Fol00} the following relations are reported
\begin{eqnarray}
\lcov\left<\phi(\tau-\lcov)\phi\right>&=&-R/(4\pi)\nonumber\\
-\left<\phi\lcov\phi\right>&=&\frac{1}{4\pi}\log\left(\frac{1}{a^2\tau}\right)+\frac{R}
{24\pi\tau}+\mathcal{O}(\tau^{-2}).
\end{eqnarray}
The density $\rho$ can now be obtained using the fact that
the metric $g_{ij}$ is conformally equivalent to the underlying metric $G_{ij}$
(up to order $1/\tau$). With the further help of equation (\ref{curvedensity}) we find as our final result\footnote{The terms of order $\tau^0$ and $\tau^{-1}$ do not depend on the coefficient $\eta$.}
\begin{eqnarray}
\label{incompress}
2\pi\rho&=&R/2-\frac{1}{2}\lcov\log\left(\frac{1}{8\pi}\log\left(\frac{1}{a^2\tau}\right)-\frac{R}{24\pi\tau}\right)\nonumber\\
&\approx&
\frac{R}{2}+\frac{1}{6\tau\log\left(1/(a^2\tau)\right)}\lcov R\nonumber\\
&=&\frac{R}{2}+\frac{1}{6\pi\tau Z}\lcov\frac{R}{2}
\end{eqnarray}
where $Z=\log(1/(a^2\tau))/(2\pi)$. 

Expression (\ref{incompress}) demonstrates, that the `gas' of zeros of an incompressible or irrotational Gaussian vector field or likewise the set of extremal points of the potential $\phi$ (random surface) behaves \textit{approximately} as the two-dimensional Coulomb gas in the high temperature phase. The latter system is well described by the Debye H\"uckel Hamiltonian for the continuous charge density $\rho$
\begin{equation}
\fl\frac{H}{T}=\frac{K_A}{2}\int\rmd A\int\rmd A'\;(2\pi\rho-R/2)_\sigma\;
G(\sigma,\sigma')
\;(2\pi\rho-R/2)_{\sigma'}+\frac{1}{2x}\int\rmd A\;\rho^2
\end{equation}
where $x$ is the fugacity of the charges. By setting $\delta H/\delta\rho=0$,
we obtain for the mean charge density
\begin{equation}
\label{dbh}
2\pi\rho=\frac{1}{1-\frac{1}{4\pi^2 K_A x}\Delta}K=K+\frac{1}{4\pi^2 K_A x}
\Delta K+\Or (x^{-2})
\end{equation}
which is equivalent to (\ref{incompress}) provided we identify $6\tau Z=4\pi K_A
x$.
The analogy  between the extremal points of $\phi$ and the Coulomb-gas is not perfect and is valid only in the high-temperature region (see \cite{Fol00,Fol01}). As already pointed out in section \ref{sec-gradient}, the scalar field $\phi$ is a nonsingular, single-valued object and must be distinguished from a field of polar angles $\theta$. The latter field (not considered here) is multivalued and has singularities which behave \textit{exactly} as Coulomb charges \cite{Kos72,Vil75}.  

For a flat geometry $G_{ij}=\delta_{ij}$, the average defect density is zero, since
\begin{equation}
\left<\partial_i\phi\partial_j\phi\right>=\frac{\delta_{ij}}{2(2\pi)^2}\int\rmd^2p\frac{1}{p^2+\tau}
\end{equation}
is also a flat metric with vanishing curvature.


\section{The two-point function}
\label{twopoint}
The calculation of the two-point function $C(r)\equiv\left<\rho(\bi{r})\rho(0)\right>$  is done here for a Euclidean $d$-dimensional geometry and a particular class of (isotropic) vector fields $u_i$ with \textit{independent} components $u_1$ and $u_2$. The correlation function
reads
\begin{equation}
\label{simplevec}
\left<u_i(\bi{r})u_j(\bi{r}')\right>=\delta_{ij}G(|\bi{r}-\bi{r}'|)
\end{equation}
equipped with an appropriate short-distance cutoff.
Examples are random waves \cite{Ber77} with $G(r)=J_0(kr)$ and the Gaussian XY-model \cite{Hal81} with $G(r)=K_0(\sqrt{\tau}r)$, where $J_0$ is zeroth order Bessel function and $K_0$ is the modified Bessel function of order zero.
Gradients of Gaussian scalar fields do not fall into the above class, since the components of the gradient $\partial_i\phi$ are not independent.
 The metric tensor and its inverse are now
\begin{eqnarray}
g_{i\alpha,j\beta}&=&\delta_{ij}\left(\delta_{\alpha\beta}G_0+c_{\alpha\beta}G(r)\right)
\nonumber\\
g^{i\alpha,j\beta}&=&\delta_{ij}\frac{\delta_{\alpha\beta}G_0-c_{\alpha\beta}G(r)}
{G_0^2-G^2(r)}
\end{eqnarray}
where $G_0=\Width+G(0)$, $c_{\alpha\beta}=1-\delta_{\alpha\beta}$, $i,j=1\ldots d$ and $\alpha,\beta=1,2$. The determinant of $g$ is $\det g=(G_0^2-G^2(r))^d$. The affine connection is
\begin{equation}
\Gamma_{k\gamma,i\alpha;m\mu}=\delta_{\alpha\gamma}\left<
\partial_iu_k(\bi{r}_\gamma)u_m(\bi{r}_\mu)\right>
=\delta_{\alpha\gamma}e_{\gamma\mu}\delta_{km}\partial_iG(r)
\end{equation}
where $\gamma$ is not summed and $e_{12}=1, e_{21}=-1, e_{11}=e_{22}=0$. We obtain the curvature tensor 
\begin{eqnarray}
R_{k\gamma,l\lambda,i\alpha,j\beta}&=&\delta_{kl}\left(\delta_{\alpha\gamma}\delta_{\beta\lambda}-\delta_{\alpha\lambda}\delta_{\beta\gamma}\right)
\left(-\partial_i\partial_jG(r)\right.\nonumber\\
&&\left.
-e_{\gamma\mu}e_{\lambda\nu}\partial_iG(r)\partial_jG(r)\frac{\delta_{\mu\nu}G_0-c_{\mu\nu}G(r)}{G_0^2-G^2(r)}\right)
\nonumber\\
&=&-\delta_{kl}e_{\alpha\beta}
e_{\gamma\lambda}\left(\partial_i\partial_jG(r)+
\partial_iG(r)\partial_jG(r)\frac{G(r)}{G_0^2-G^2(r)}\right)\nonumber\\
&=&-\sqrt{G_0^2-G^2(r)}\,\delta_{kl}e_{\alpha\beta}
e_{\gamma\lambda}\partial_i\partial_jK(r)
\end{eqnarray}
where $K(r)=\arcsin(G(r)/G_0)$ and the indices $\gamma,\lambda$ are not summed. The two-point correlation function $C(r)=\left<\rho(\bi{r})\rho(0)\right>$ now reads
\begin{eqnarray}
(2\pi)^dC(r)
&=&(G_0^2-G^2(r))^{-d/2}\int(\rmd\eta\rmd\theta)^{2d}\nonumber\\
&&\times\exp\left(\theta_k^{1}\theta_k^{2}\eta_i^{1}\eta_j^{2}\sqrt{G_0^2-G^2(r)}\,\partial_i\partial_jK(r)\right)\nonumber\\
&=&(-1)^dd!\int (\rmd\eta^2\rmd\eta^1)^d\exp\left(\eta^1_i\eta^2_j\partial_i\partial_jK(r)\right)\nonumber\\
&=&(-1)^dd!\det(\partial_i\partial_jK(r))
\end{eqnarray}
since the term $(\ldots)^d/d!$ of the exponential is the only one which contributes to the result. The square root $\sqrt{G_0^2-G^2(r)}$ is therefore cancelled by the prefactor. The integration over the $\theta$-variables yields an extra factor $(-1)^dd!$.
Denoting $\partial(\ldots)/\partial r=(\ldots)'$ we end up with
\begin{equation}
\label{corrfinal}
(2\pi)^dC(r)=(-1)^dd! K(r)'' \left(K(r)'/r\right)^{d-1}.
\end{equation}
This result was already obtained by direct calculations \cite{Sai01,Hal81,Ber00,Liu92} for the limiting case $\Width\rightarrow 0$. 
The correlation function $C(r)$ is a functional of the ratio $G(r)/G_0$ and is therefore independent of the scale $T$ of the field $u_i$ (in the limit of a sharp delta function $\Width\rightarrow 0$).

\subsection{Singular behaviour of correlations for coinciding points and neutrality}
\label{sumrule}
Each vector field placed on a closed surface is subject to the topological constraint (\ref{topol}), implying a neutrality condition for correlation functions
\begin{equation}
\int\rmd^2\sigma\sqrt{\det G}\left<\rho(\sigma)\rho(\sigma')\right>=\langle\rho(\sigma')\sum_\alpha q_\alpha\rangle
=2(1-\gamma)\left<\rho(\sigma')\right>,
\end{equation}
where $G_{ij}(\sigma)$ is the metric tensor of the surface. Analogue constraints apply to higher order correlation functions.
To find a neutrality condition for vector fields on an \textit{infinite} planar geometry, we note that the density $\rho$ can be written as a divergence (for a finite width $\Width$)
\begin{eqnarray}
\rho&=&e_{i_1,\ldots,i_d}\prod_k\partial_{i_k}u_k\,(2\pi\Width)^{-1/2}\exp
\left(-(u_k)^2/(2\Width)\right)\nonumber\\
&=&e_{i_1,\ldots,i_d}\prod_k\partial_{i_k}\erf(u_k/\sqrt{\Width})\nonumber\\
&=&\partial_i\left(e_{i,i_2,\ldots,i_d}\erf(u_1/\sqrt{\Width})\prod_{k=2}^d
\partial_{i_k}\erf(u_k/\sqrt{\Width})\right)
\end{eqnarray}
where $\erf(x)=(2\pi)^{-1/2}\int_0^x\rmd s\exp(-s^2/2)$ is the error function and $e_{i_1,\ldots,i_d}$ is the antisymmetric unit tensor in $d$ dimensions. Then the total charge of a vector field is zero, i.e.
\begin{equation}
\label{planarsr}
\int\rmd^2r\left<\rho(\bi{r})\rho(\bi{r}')\right>=0
\end{equation}
provided that the correlations $\left<\erf(u_i(0))\erf(u_j(\bi{r}))\right>\sim \left<u_i(0)u_j(\bi{r})\right>$ decay fast enough (at least   $\sim r^{-1/2}$)  for $r\rightarrow\infty$.
On the other hand, the correlation function should behave like a delta-function for $\bi{r}\rightarrow\bi{r}'$ and $\Width\rightarrow 0$ since
\begin{eqnarray}
\label{neutral}
\lefteqn{\left<\rho(\bi{r})\rho(\bi{r}')\right>=\sum_{\alpha\beta}
\left<q_\alpha\delta^2(\bi{r}-\bi{r}_\alpha)
\,q_\beta\delta^2(\bi{r}'-\bi{r}_\beta)\right>}\nonumber\\
&=&\sum_\alpha\left<\delta^2(\bi{r}-\bi{r}_\alpha)
\delta^2(\bi{r}'-\bi{r}_\alpha)\right>
+\sum_{\alpha\neq\beta}
\left<q_\alpha\delta^2(\bi{r}-\bi{r}_\alpha)
\,q_\beta\delta^2(\bi{r}'-\bi{r}_\beta)\right>\nonumber\\
&=&\delta^2(\bi{r}-\bi{r}')
\sum_\alpha\left<\delta^2(\bi{r}-\bi{r}_\alpha)
\right>+\sum_{\alpha\neq\beta}
\left<q_\alpha\delta^2(\bi{r}-\bi{r}_\alpha)
\,q_\beta\delta^2(\bi{r}'-\bi{r}_\beta)\right>\nonumber\\
&=&n_0(\bi{r})\delta^2(\bi{r}-\bi{r}')+\sum_{\alpha\neq\beta}
\left<q_\alpha\delta^2(\bi{r}-\bi{r}_\alpha)
\,q_\beta\delta^2(\bi{r}'-\bi{r}_\beta)\right>
\end{eqnarray}
where $n_0=\left<|\det(\partial_iu_j)|\delta^2(\bi{u})\right>$ is the absolute density of zeros \cite{Hal81}. We consider now a planar geometry and  vector fields with an isotropic 
distribution. Then $n_0$ is spatially constant and both $\left<\rho(\bi{r})\rho(\bi{r}')\right>$ and  $\sum_{\alpha\neq\beta}
\left<q_\alpha\delta^2(\bi{r}-\bi{r}_\alpha)
\,q_\beta\delta^2(\bi{r}'-\bi{r}_\beta)\right>$ are functions of $|\bi{r}-\bi{r}'|$ only. We integrate equation (\ref{neutral}) over $\bi{r}'$  and obtain
(see also \cite{Hal81,Liu92,Fre98,Ber00}) 
\begin{equation}
0=n_0+\int_{|\bi{r}|>a}\rmd^2rC(r)
\end{equation}
where we have used equation (\ref{planarsr}). 
The integral extends over two-dimensional space with the exception of an infinitesimal small disk (radius $a\rightarrow 0$) at the origin.
For $d=2$  and vector fields with correlation function (\ref{simplevec})
we have found (see equation (\ref{corrfinal}) in the preceding section)
\begin{equation}
\label{corrfu}
C(r)=\frac{\left((K'(r))^2\right)'}{(2\pi)^2r}.
\end{equation}
The correlation function of the vector field $u$ must be  
regularized at short distances and can therefore be expanded for small radii
$G(r)=G(0)+r^2G''(0)/2+\ldots\,(G''(0)<0)$, which implies (for small $\Width$)
\begin{eqnarray}
K(r)&=&\arcsin(G(r)/G_0)=\frac{\pi}{2}-\left(\frac{2\Width}{G_0}+\frac{-G''(0)}{G_0}r^2\right)^{1/2}+\ldots\nonumber\\
&=&\frac{\pi}{2}-\sqrt{\frac{-G''(0)}{G_0}}\sqrt{a^2+r^2}+\ldots
\end{eqnarray} where $a^2\equiv2\Width/(-G''(0))$.
We obtain for $\Width\rightarrow 0$ with $K'(r)\rightarrow 0$ for $r\rightarrow\infty$
\begin{equation}
-n_0=2\pi\int_{0_+}^\infty r\rmd r C(r)=\frac{1}{2\pi}\left.\left(K'(r)\right)^2
\right|_{0_+}^\infty=\frac{G''(0)}{2\pi G(0)}.
\end{equation}
The function $K(r)$ has for $\Width\rightarrow 0$ a conical singularity at the origin. The correlation function $C(r)\propto\det(\partial_i\partial_jK(r))$ can now be interpreted as the Gaussian curvature of that cone, which is in fact concentrated at its tip. 
Using equation (\ref{corrfu}), we find the anticipated short-distance behaviour of the two-point correlation function $C(r)$
\begin{equation}
C(r)\sim\left(\frac{-G''(0)}{G_0}\right)\frac{2a^2}{(2\pi)^2(r^2+a^2)^2}\sim
\frac{-G''(0)}{2\pi G(0)}\delta^2(\bi{r})=n_0\delta^2(\bi{r}).
\end{equation}

\section{Conclusions}

We have proposed a scheme to evaluate correlation functions of the (charge-) density of zeros $\rho$ of random vector fields $u_i$ with a Gaussian distribution. 
The zeros carry a charge $q=\textrm{sign} J=\pm 1$, where $J=\det(\partial_i u_j)$ is the Jacobian of the vector field right at the zero. In two dimensions for instance, positive zeros resemble sources, sinks or vortices, while negative ones have a saddle-like flow.
We could show that the correlation functions $\left<\rho(\bi{r}_1)\ldots
\rho(\bi{r}_f)\right>$ can be expressed through the curvature of an abstract
$f\times d$-dimensional Riemann-Cartan manifold ($d$ is the dimension of both the vector field and the embedding space). In the case of gradient fields $u_i=\partial_i\phi$, the correlation functions are the total curvature of a certain $f\times d$-dimensional  Riemannian manifold. 
As an application, we have calculated the mean charge density for various vector fields on curved two-dimensional surfaces. Furthermore, we derived the two-point function for isotropic vector fields with independent components. We checked the neutrality sum rule, which follows from the topological constraint of the total charge and obtained in addition the short-distance behaviour of the two-point correlation function.
We have seen that the zeros of Gaussian vector fields are a model for the two-dimensional Coulomb gas in the high-temperature region. This opens a convenient route to study these systems in curved or bounded geometries. The geometrical view helps to calculate one- and two-point correlation functions and could also be rather useful to `tame' the higher order correlation functions of Gaussian zeros.

The present work demonstrates that the signed zeros of vector fields with a Gaussian distribution are not only sensitive to curvature, the correlation functions of the charge density \textit{are} the curvature of an abstract manifold, which is closely related to the embedding manifold.


\ack

I would like to thank H K Janssen for encouraging discussions.
This work has been supported by the Deutsche Forschungsgemeinschaft
under SFB 237.

\end{document}